\documentclass{sigma}

\def \arccosh{\mathop{\rm arccosh}\nolimits}

\def \csch{\mathop{\rm csch}\nolimits}

\begin{document}

\renewcommand{\PaperNumber}{***}

%\FirstPageHeading

\ShortArticleName{On the Quasi-Exact Solvability of the Confluent Heun Equation}

\ArticleName{On the Quasi-Exact Solvability of the Confluent Heun Equation}

% Names of the authors for the title of the paper
\Author{M.A. Gonz\'alez Le\'on$^{\dag}$, J. Mateos Guilarte$^{\ddag}$, A. Moreno Mosquera$^{\S}$ and M. de la Torre Mayado$^{\ddag}$}

\AuthorNameForHeading{M.A. Gonz\'alez Le\'on, J. Mateos Guilarte, A. Moreno Mosquera and M. de la Torre Mayado}

\Address{$^{\dag}$~Departamento de Matem\'atica Aplicada and IUFFyM, Universidad de Salamanca, Spain} % Address of First Author
\EmailD{magleon@usal.es} % E-mail address of First Author
\URLaddressD{http://campus.usal.es/\~{}mpg} %URL address of First Author

% Address of Second Author
\Address{$^{\ddag}$~Departamento de F\'{\i}sica Fundamental and IUFFyM, Universidad de Salamanca, Spain}
\EmailD{guilarte@usal.es, marina@usal.es} % E-mail address of Second Author

\Address{$^{\S}$~Facultad Tecnol\'ogica, Universidad Distrital Francisco Jos\'e de Caldas, Bogot\'a, Colombia} % Address of First Author
\EmailD{asmorenomosquera@gmail.com} % E-mail address of First Author

%\ArticleDates{Received ???, in final form ????; Published online ????}

\Abstract{
It is shown that the Confluent Heun Equation (CHEq) reduces for certain conditions of the parameters to a particular class of Quasi-Exactly Solvable models, associated with the Lie algebra $sl (2,{\mathbb R})$. As a consequence it is possible to find a set of polynomial solutions of this quasi-exactly solvable version of the CHEq. These finite solutions encompass previously known polynomial solutions of the Generalized Spheroidal Equation, Razavy Eq., Whittaker-Hill Eq., etc. The analysis is applied to obtain and describe special eigen-functions of the quantum Hamiltonian of two fixed Coulombian centers in two and three dimensions.}

%\pacs{02.30.Hq, 03.65.Fd, 03.65.Ge}
%\noindent{\it Keywords}: Confluent Heun equation, Quasi-Exactly Solvable systems, Two Coulombian centers problem.

%Generalized Spheroidal equation, Razavy and Whittaker-Hill equations.

\section{Introduction}

The Confluent Heun Equation (CHEq) is the result of the simplest case of confluence in the Heun Equation, a Fuchsian equation with four regular singular points, which in standard form \cite{Slavyanov} are located at $z=0,1, a$ and $\infty$. The confluence of the last two singularities into an irregular singular point at $z=\infty$ gives rise to the CHEq. There exist other forms of confluence for the Heun equation: Double confluent, Bi-confluent and Tri-confluent Heun equations \cite{Slavyanov, Ronveaux}. In recent years a vast body of literature about Heun equations and their different confluent forms has been published. Applications to physics and the other natural sciences are very numerous, see \cite{Slavyanov, Ronveaux, Hortacsu} and references therein.

The Confluent Heun equation encompasses as particular cases a plethora of very well known equations in Mathematical Physics, such as the Spheroidal and Generalized Spheroidal Wave equations, the Whittaker-Hill and the Razavy Equations, or the Mathieu equation. In Section 2 we shall show how all these equations arise as special cases of the CHEq. It has been observed that under certain conditions in the parameter space polynomial solutions exist in all the above equations; see References \cite{Wilson} and \cite{Demkov} to find finite solutions of the Generalized Spheroidal Equation. Moreover, in References \cite{Artemio} and \cite{GGTCM} polynomial solutions of the Razavy and Whittaker-Hill Equations are described, whereas in \cite{Ronveaux} finite solutions to the CHEq itself are discussed.

In this work we analyze the Confluent Heun equation in the context of one-dimensional Quasi-Exactly Solvable (QES) models \cite{Turbiner, Shifman}, i.e. the class of spectral problems where an arbitrary part of the eigenvalues and eigenfunctions can be found algebraically, but not the whole spectrum. Following the notation introduced in \cite{Olver} for the QES models that admit normalizable solutions, we shall prove in Section 3 that the CHEq is equivalent to the most general case of QES systems of type II. According to this, an arbitrary number of polynomials solutions of the CHEq can be found, and these solutions encompass as particular cases all the previously known finite solutions of some of the Mathematical Physics equations listed above. The general procedure of searching for polynomial solutions is developed in Section 4.

The Generalized Spheroidal equation (GSEq) appears in a wide set of problems in different areas of physics as general relativity \cite{Leaver, Leaver2, Figueiredo}, molecular quantum mechanics \cite{Wilson, Demkov, Leaver2, Ponomarev}, polymer physics \cite{Vincenzi} or condensed matter physics \cite{Shan}. In Section 5 we analyze the quantum problem of two fixed Coulombian centers in the space ${\mathbb R}^3$ \cite{Wilson, Turbiner3, Turbiner2, Ponomarev}. The corresponding Schr\"odinger equation splits into two different GS equations, one for the range $z\in (1,\infty)$, and the second one for $z\in (-1,1)$. Forcing quasi-exact solvability in both equations is equivalent to find in the energy spectrum the structure of the hydrogenoid atoms levels. Moreover, compatibility of the separation constant determines the distance between the centers where these eigenfunctions arise. In this context, the solutions found by Demkov in \cite{Demkov} are calculated from an algebraic point of view. Finally, we analyze the analogue two-dimensional problem \cite{ GGTCM, Turbiner2, GMT2007, GMT2012} where the r\^ole of the GSEq is played by the Razavy and Whittaker-Hill equations. There appear different possibilities to identify these equations as particular cases of CHEq, that lead to an interesting structure of the so-called Demkov solutions in the two dimensional case.

\section{The Confluent Heun and some descendent equations}

In standard form the Confluent Heun equation is written \cite{DLMF} as:
\begin{equation}
z(z-1)\, u''(z) + \left( \epsilon z(z-1)+\gamma (z-1)+\delta z\right)  u'(z)+ (\alpha z-q) u(z) = 0\label{CHE} \, .
\end{equation}
It is a linear ODE of second-order that depends on five real parameters: $\alpha, \gamma, \delta, \epsilon$ and $q$. Equation (\ref{CHE}) has two regular singular points located at $z=0$ and $z=1$, with characteristic exponents: $(0,1-\gamma)$ and $(0, 1-\delta)$ respectively. This equation also has an irregular singular point at infinity as the result of the confluence of the other two regular singular points of the Heun Equation.  It is usual to rewrite equation (\ref{CHE}) in a more symmetric way, locating the regular singular points at $z=\pm 1$, by re-scaling and translating the independent variable $z\to 2z-1$, to obtain
\begin{equation}
\left( z^2-1\right)u''(z)+ \left( \frac{\epsilon}{2} (z^2-1)+ \gamma (z-1)+\delta(z+1)\right) u'(z)+ \left( \frac{\alpha}{2} (z+1)-q\right)u(z)  = 0 \, . \label{CHEs}
\end{equation}
Alternatively, the CHEq (\ref{CHEs}) can be interpreted as the spectral problem
\begin{equation*}
D\, u(z)\, =\, q\, u(z) \, ,
\end{equation*}
associated with the differential operator
\begin{equation}
D =\left( z^2-1\right) \frac{d^2}{dz^2} + \left( \frac{\epsilon}{2} (z^2-1) + \gamma (z-1)+\delta(z+1)\right) \frac{d}{dz}+  \frac{\alpha}{2} (z+1) \, ,\label{CHEsoperator}
\end{equation}
$q$ being the corresponding eigenvalue. From this point of view, it is natural to write CHEq as a Sturm-Liouville equation:
\begin{equation}
\frac{1}{\omega(z)}\, \frac{d}{dz} \left( (z+1)^\gamma (z-1)^\delta \, e^{\frac{\epsilon}{2}\, z} \, \frac{du(z)}{dz}\right) \, +\, \frac{\alpha}{2} (z+1) u(z)\, =\, q \, u(z)\label{SL}
\end{equation}
with weight function
\[
\omega(z)\, =\,  (z+1)^{\gamma-1} (z-1)^{\delta-1} \, e^{\frac{\epsilon}{2}\, z}\, .
\]

The CHEq (\ref{CHEs}) can be transformed by using different changes of the dependent variable into several very well known equations in Mathematical Physics. First we shall consider the change of variable:
\[
u(z)=(z+1)^{\frac{1-\gamma}{2}} \, (z-1)^{\frac{1-\delta}{2}}\, e^{-\frac{\epsilon z}{4}}\, v(z)\   .
\]
In terms of the new variable $v(z)$, equation (\ref{CHEs}) reads:
\begin{equation}
\frac{d}{dz} \left( (z^2-1) \frac{dv(z)}{dz} \right)\, +\, \left( \, A\, z^2\, +\, B\, z\, +\, C\, +\, \frac{a\, z+b}{z^2-1}\right) \, v(z)\, =\, 0\label{CHEs1}\, ,
 \end{equation}
where the new parameters are defined in terms of the old ones:
\begin{eqnarray*}
&& A = -\frac{\epsilon^2}{16} \quad , \quad B= \frac{\alpha}{2} \, -\, \frac{\epsilon}{4} \, (\gamma+\delta) \\ &&C= \frac{\epsilon^2}{16}+\frac{\epsilon}{4} \, (\gamma-\delta)-\frac{\gamma+\delta}{4} \, (\gamma+\delta-2)\, +\frac{\alpha}{2} -q \\ &&  a= \delta \, ( 1-\frac{\delta}{2}) -  \gamma \, ( 1-\frac{\gamma}{2})\quad , \quad b= \delta \, ( 1-\frac{\delta}{2}) +\gamma \, ( 1-\frac{\gamma}{2}) \, -\, 1 \, .
\end{eqnarray*}

Equation (\ref{CHEs1}), i.e., the Confluent Heun Equation written for the function $v(z)$, reduces to the Generalized or Coulomb Spheroidal Equation \cite{DLMF} for the special value $a=0$. This restriction requires the following relations between the original parameters: (1) either $\gamma=\delta$, (2), or $\gamma=2-\delta$. The Confluent Heun Equation transforms into the Generalized Spheroidal Equation in two different ways:

\noindent $\bullet$ $\gamma=\delta$. The corresponding change of variable is:
\[
u(z)\, =\, (z^2-1)^{\frac{1-\delta}{2}} \, e^{-\frac{\epsilon z}{4}}\,  v(z) \, ,
\]
whereas the parameters influenced by the restriction become: $B = \frac{1}{2} \left( \alpha-\epsilon\, \delta\right)$, $C= \frac{\epsilon^2}{16}+\frac{\alpha}{2} -q+\delta (1-\delta) $, $b= -(1-\delta)^2$.

\noindent $\bullet$ $\gamma=2-\delta$. The appropriate change of variable is:
\[
u(z)\, =\, \left(\frac{z-1}{z+1}\right)^{\frac{1-\delta}{2}} \, e^{-\frac{\epsilon z}{4}}\,  v(z)
 \]
and the restricted parameters are:  $B= \frac{1}{2} \left(\alpha - \epsilon\right)$, $C= \frac{\epsilon^2}{16}+\frac{\epsilon (1-\delta)}{2} \, +\frac{\alpha}{2} -q$, $b= -(1-\delta)^2$.

The Spheroidal Wave Equation \cite{DLMF} is simply the particular $a=B=0$ case of equation (\ref{CHEs1}), i.e.,
\begin{equation}
\frac{d}{dz} \left( (z^2-1) \frac{dv(z)}{dz} \right)\, +\, \left( \, A\, z^2\, +\, C\, +\, \frac{b}{z^2-1}\right) \, v(z)\, =\, 0\label{SEq1}\, .
\end{equation}
Equivalently, one descends from equation (\ref{CHEs1}) to equation (\ref{SEq1}) by imposing the relations between the original parameters: (1) $\alpha=\epsilon \delta$ if $\gamma=\delta$, and, (2) $\alpha=\epsilon$ for $\gamma=2-\delta$. Finally, the special value $A=0$, i.e. $\epsilon =0$ in the original set of parameters, leads from equation (\ref{SEq1}) to the Associated Legendre Equation, which in turn reduces to the Legendre Equation for $b=0$, i.e., $\delta=\gamma=1$.

It is possible to choose another descending path from the CHEq to a second set of important equations in Mathematical Physics taking the change of variable:
\[
u(z)=(z+1)^{\frac{1-2\gamma}{4}} \, (z-1)^{\frac{1-2\delta}{4}}\, e^{-\frac{\epsilon z}{4}}\, w(z)\   .
\]
Equation (\ref{CHEs}) thus becomes:
\begin{equation}
\left( z^2-1\right) w''(z) +  z w'(z)  + \left( A  z^2 + B z + C-\frac{1}{4} + \frac{a z+b+\frac{1}{4}}{z^2-1}\right)  w(z) = 0\, .\label{CHEs2}
 \end{equation}
If $a=0$ and $b=-\frac{1}{4}$, equation (\ref{CHEs2}) is the algebraic form of the Razavy and Whittaker-Hill Equations
\begin{equation}
(z^2-1)\, w''(z) \, +\, z\, w'(z)\, +\, \left( A\, z^2+B \, z+ C-\frac{1}{4}\right)  \, w(z)\,  =\, 0\label{RWH1}\, ,
\end{equation}
which, in the context of the original parameters of the CHEq, comes from four concrete combinations of parameters: 1) $\gamma=\delta=\frac{3}{2}$, 2) $\gamma=\delta=\frac{1}{2}$, 3) $\gamma=\frac{1}{2}\, ,\ \delta=\frac{3}{2}$, 4) $\gamma=\frac{3}{2}\, ,\ \delta=\frac{1}{2}$.
Finally, adding the additional condition $B=0$ we obtain equation
\begin{equation}
(z^2-1)\, w''(z) \, +\, z\, w'(z)\, +\, \left( A\, z^2+ C-\frac{1}{4}\right)  \, w(z)\,  =\, 0\label{Mat1}\, ,
\end{equation}
which is the algebraic form of the Trigonometric and Hyperbolic Mathieu equations.

\section{Quasi-Exact Solvability and the Confluent Heun Equation}

Quasi-Exactly-Solvable (QES) systems are very interesting spectral problems characterized by the fact that part of the eigenvalues and eigenfunctions, but not the whole spectrum, can be found algebraically. In particular, an important class of QES systems has a dynamics characterized by a Hamiltonian that is an element of the enveloping algebra of a finite-dimensional Lie algebra of differential operators, which in turn admits a finite dimensional invariant module of smooth functions. After the pioneer works of Turbiner, Ushveridze and Shifman, \cite{Turbiner, Shifman, Ushveridze, TurShifman}, a growing body of literature addressing QES systems has appeared over the past twenty years.

One delicate point is the analysis of the normalizability of the wave functions. In \cite{Olver} the necessary and sufficient conditions for the normalizability of the algebraic part of the spectrum were determined. In particular, QES systems that admit normalizable algebraic eigenfunctions have been classified in several canonical forms. Following the notation used in \cite{Olver} for a generic second-order QES spectral problem, ${\cal H} f(z)\, =\, \lambda f(z)$, the Hamiltonian operator ${\cal H}$ is written, possibly after an adequate ``gauge" transformation, as a quadratic combination:
\begin{eqnarray}
{\cal H}& =&  \, c_{++} \, (J^+)^2\,+\, c_{00}\, (J^0)^2\, +\, c_{--}\, (J^-)^2 \, +\,  c_{+0} \, \left( J^+J^0 +  J^0J^+\right)\, + \nonumber \\ &&+ c_{+-} \left( J^+J^- +  J^-J^+\right)+ c_{0-} \left( J^0J^- +  J^-J^0\right) + c_+ J^+ +  c_0 J^0 + c_- J^-\label{ham}
\end{eqnarray}
of $J^-$, $J^+$ and $J^0$, the generators of the $sl (2,{\mathbb R})$ Lie algebra:
\[
J^-\, =\, \frac{d}{dz}\  ,\quad J^0\, =\, z\, \frac{d}{dz}\, -\, \frac{n}{2}\  ,\quad J^+\, =\, z^2\, \frac{d}{dz}\, -\, nz \, \, .
\]
Here, $n$ is a non-negative integer that determines the dimension of the invariant module and the three differential operators close the algebra:
\[
\left[ J^0,J^+ \right]\, =\, J^+\  ,\quad \left[ J^0,J^- \right]\, =\, -J^-\  ,\quad \left[ J^+,J^- \right]\, =\, -2 J^0\, .
\]
Developing the quadratic combination (\ref{ham}) for a given value of $n$, and equating the expansion in powers of $\frac{d}{dz}$ up to second order to the corresponding coefficients of the operator (\ref{CHEsoperator}), it is not difficult to check that $D$ can be written in such a form (\ref{ham}) if and only if the parameters $\alpha$ and $\epsilon$ satisfy the relationship:
\begin{equation}
\alpha\, =\, -n\, \epsilon\label{QEScondition}\, .
\end{equation}
Moreover, if this is the case, $\frac{\alpha}{\varepsilon}$ is a negative integer, $D$ is a QES operator of the form:
\begin{eqnarray}
D&=& 2 \frac{n+\gamma +\delta -\epsilon }{n+2}  J^0 J^0 -  J^-J^-  + \frac{2-n-2(\gamma+\delta-\epsilon)}{2(n+2)} \left(  J^+J^-+J^-J^+  \right)\nonumber  \\ && +\, \frac{\epsilon}{2}\, J^+\, +\, \left( \gamma +\delta+ n -1\right) \, J^0\, +\, \left(\delta-\gamma-\frac{\epsilon}{2}\right) \, J^-\label{QESCHE1}
\end{eqnarray}
Thus, the CHEq is a Quasi-Exactly-Solvable system if and only if (\ref{QEScondition}) is satisfied, for some $n\in {\mathbb N}$.

According to Theorem 7 in \cite{Olver}, any quasi-exactly solvable operator of the general form
\[
-{\cal H}\, =\, P_2(z) \, \frac{d^2}{dz^2} \, +\, P_1(z)\, \frac{d}{dz}\, +\, P_0(z)\, ,
\]
where $P_2(z)>0$ on an interval $I\subset {\mathbb R}$, can be written in Schr\"odinger form:
\[
\mu(z)\, \cdot \, {\cal H} \, \cdot \, \frac{1}{\mu(z)}\, =\, -\frac{d^2}{dx^2}+V(x) \, \, .
\]
In this equation $x$ is a new variable determined through the change $z=\zeta(x)$, such that $\left( \frac{dz}{dx}\right)^2\, =\, P_2(z)$, whereas  the ``gauge" factor $\mu(z)$ is more complicated:
\[
\mu(z)\, =\, \left| P_2(z)\right|^{-\frac{n}{4}} \, {\rm exp}\left\{ \int^z \frac{P_1(y)+\frac{n-1}{2} P'_2(y)}{2\, P_2(y)} \, dy \right\}\, .
\]
By construction, if $\chi(z)$ is an eigenfunction of the original operator ${\cal H}$, the function $\psi(x)\, =\, \mu(\zeta(x))\, \chi(\zeta(x))$ will be the corresponding eigenfunction of the Schr\"odinger operator with the same eigenvalue.

Using this notation, and restricting the $z$ coordinate to the range $|z|>1$, in order to satisfy the hypothesis of the theorem the CHEq operator (\ref{QESCHE1}) can be written as the Schr\"odinger operator
\begin{eqnarray}
\mu(z)\cdot{\cal H} \cdot \frac{1}{\mu(z)} &=& -\frac{d^2}{dx^2} - A \cosh^2  x - B \cosh  x - a \coth  x\, \csch  x\nonumber \\ &&- \left( b+\frac{1}{4}\right)\csch^2  x-  C+\frac{1}{4}\label{SchrCHEq}
\end{eqnarray}
where $x\, =\, \arccosh z$, and $\mu(z)\, =\, (z-1)^{\frac{2\delta-1}{4}} (z+1)^{\frac{2\gamma-1}{4}}\, e^{\frac{\epsilon z}{4}}$. This is remarkable: the gauge factor is precisely the factor that determines the change of variable that converts CHEq into equation (\ref{CHEs2}). The parameters $A$, $B$, $C$, $a$ and $b$ in the differential operator in (\ref{SchrCHEq}) are exactly the same as those appearing in the differential operator in (\ref{CHEs1}) after imposing the QES condition $\alpha\, =\, -n\epsilon$. Thus, $B$ and $C$ become:
\begin{eqnarray*}
B&=& -\frac{\epsilon}{4} \, \left(2n+\gamma+\delta\right)\\ C&=& \frac{1}{16} \left( \epsilon-2\delta-2\gamma\right)^2+\frac{\gamma}{2}(1-\gamma)+\frac{\delta}{2}(1-\delta)-q-\frac{n\epsilon}{2}\, .
\end{eqnarray*}
Up to the constant term, $C-\frac{1}{4}$, the operator (\ref{SchrCHEq}) exhibits the canonical form of any QES spectral problem of type II in the classification achieved in \cite{Olver}. We conclude that the Confluent Heun Equation encompasses a broad class of QES one-dimensional spectral problems.

\section{Polynomial solutions}

Having proved the quasi-exact solvability of the CHEq if $\alpha=-n\, \epsilon$, let us determine the invariant module of polynomial solutions associated with an arbitrary value of the non-negative integer $n$. In this QES version, the CHEq is written as:
\begin{equation}
\left( z^2-1\right)\, u''(z)\, +\, \left( \frac{\epsilon}{2} (z^2-1)\, +\, \gamma (z-1)+\delta(z+1)\right)\, u'(z)\,   +\, \left( \frac{-n\epsilon}{2} (z+1)-q\right)\, u(z)\, =\, 0 \label{CHEQES}
\end{equation}
We shall consider one of the Frobenius solutions associated with the $z=-1$ regular singular point
\begin{equation}
u(z)\, =\,   \sum_{k=0}^\infty\, \frac{(-1)^k \,  {\cal P}_k(q)}{2^{k}\,k!\, (\gamma)_k } \, \,  (z+1)^k \label{frob1}\, ,
\end{equation}
where $(\gamma)_k=\frac{\Gamma(\gamma+k)}{\Gamma(\gamma)}=\gamma (\gamma+1)\dots (\gamma+k-1)$ denotes the Pochhammer symbol, the special form of the coefficients of the series (\ref{frob1}) has been chosen in such a way that ${\cal P}_k(q)$ will be a monic polynomial of degree $k$ in the parameter $q$ and the zero-th-order polynomial is normalized to $1$: ${\cal P}_0(q)=1$. Substitution of (\ref{frob1}) in equation (\ref{CHEQES}) leads to ${\cal P}_1(q) \, =\, q$, whereas (\ref{frob1}) is a power series solution of (\ref{CHEQES}) if the following three-term recurrence between the polynomials for $k\geq 1$,
\begin{equation}
{\cal P}_{k+1}(q) = \left( q-k(M+k-1)\right)  {\cal P}_k(q)- k \epsilon  (n-k+1)(\gamma+k-1)  {\cal P}_{k-1}(q)\  ,\label{recurrence}
\end{equation}
holds. The notation $M=\delta+\gamma-\epsilon$ has been introduced in (\ref{recurrence}) in order to simplify the expressions. The recurrence relation (\ref{recurrence}) can easily be solved in low orders such that the following coefficients are determined up to fourth order:
\begin{eqnarray*}
{\cal P}_1(q)&=& q\\ {\cal P}_2(q)&=& q^2\, -\, M\, q\, -n\gamma \epsilon \\ {\cal P}_3(q)&=& q^3\, -\, (3M+2)\, q^2\, +\, \left( 2M(M+1)-\epsilon(\gamma(3n-2)+2(n-1))\right) q\\ && +2n\epsilon\gamma (M+1)\\ {\cal P}_4(q)&=& q^4-2(3M+4) q^3+\left( 11 M^2+26 M +12  -2 \epsilon (4n-7 \right. \\ && \left. \, + \gamma(3n-4))\right) q^2 +\, (-6M(M+1)(M+2)  \\ && +2\epsilon\, \left((M+1)\, (3(2n-3) +\gamma(7n-6)) +3(n\gamma+1))\right) q \\ && -\, 3n\gamma\epsilon\left( 2 (M+1)(M+2)-(n-2)(\gamma+2)\epsilon\right)\, .
\end{eqnarray*}
Looking at the ${\cal P}_{k-1}$-coefficient in the recurrence (\ref{recurrence}), it is clear that it is null if $k=n+1$: in this case the recurrence simplifies to a two-term relation:
\[
{\cal P}_{n+2}(q)\, =\, \left( q-(n+1)(M+n)\right) \, {\cal P}_{n+1}(q)\, .
\]
In this situation, if $q$ is chosen as one of the $n+1$ roots of ${\cal P}_{n+1}(q)$, $q_j$, $j=1,\dots,n+1$, all the higher order polynomials are also zero:
\[
0={\cal P}_{n+1}(q_j)={\cal P}_{n+2}(q_j)={\cal P}_{n+3}(q_j)=\dots\qquad \forall j=1,\dots, n+1 \, \, ,
\]
i.e., the series truncates to a polynomial solution of degree $n$:
\begin{equation}
u_{n,j}(z) =   1 -  \frac{q_j}{2\gamma} \,  (z+1) + \frac{q_j^2- M q_j -n\gamma \epsilon}{8 \gamma (\gamma+1) } \,  (z+1)^2 +\, \dots \, +  \frac{(-1)^n \,  {\cal P}_n(q_j)}{2^{n}\, n!\, (\gamma)_n } \, \,  (z+1)^n\label{truncatedsol}\, .
\end{equation}
It is not difficult to prove that the polynomial solutions (\ref{truncatedsol}) of (\ref{CHEQES}), $u_{n,j}(z)$, are independent of the choice of the regular singular point to apply the Frobenius method of finding power series solutions by solving/truncating the corresponding three-term recurrence. For a given value of $n$, and a fixed root $q_j$ of ${\cal P}_{n+1}(q)$, the polynomial solution $u_{n,j}(z)$  is unique up to multiplicative constants and independent of the choice of the expansion point.

The polynomials $u_{n,j}(z)$ exhibit orthogonality properties derived from the general Sturm-Liouville Theory. Assuming boundary conditions in equation (\ref{SL}) and normal assumptions in the coefficients in such a way that the Sturm-Liouville problem is standard, the polynomials $u_{n,j}(z)$ exhibit the property of double orthogonality:
\[
\int_{-1}^1 u_{n,j}(z) u_{n,l}(z) \omega(z)\, dz\, =\, \int_{1}^\infty u_{n,j}(z) u_{n,l}(z) \omega(z)\, dz\, =\, 0\, \quad,\qquad n\neq l\, .
\]
Furthermore, the eigenvalues $q_1,\dots q_{n+1}$ are single and real. If they are ordered: $q_1<q_2<\dots <q_{n+1}$, then the corresponding polynomials $u_{n,j}(z)$, $j=1,\dots,n+1$ will have $j-1$ zeroes in $(-1,1)$ and $n+1-j$ zeroes in $(1,\infty)$, see Reference \cite{Ronveaux}.

The set of polynomials $\left\{\, {\cal P}_k\right\}_{k\in {\mathbb N}}$ in itself has interesting orthogonality properties. Applying general results about orthogonal polynomials associated with QES systems, see \cite{Finkel} and references therein for details, the three-term recurrence (\ref{recurrence}), together with the truncation to a two-term one if $k=n+1$, ensures that the surviving polynomials $\left\{\, {\cal P}_k\right\}_{k\in {\mathbb N}}$ form a weakly orthogonal family. Moreover, there exists an associated moment functional ${\cal L}$ that acts in the space of complex polynomials ${\mathbb C}[q]$ in such a way that ${\cal P}_k(q)$ are orthogonal under the action of ${\cal L}$:
\[
{\cal L}\left( {\cal P}_k\, {\cal P}_l \right) \, =\, \nu_k\, \delta_{kl}\quad,\qquad k,l\in {\mathbb N}\, ,
\]
where $\delta_{kl}$ denotes the Kronecker symbol. The coefficient $\nu_k$ plays the r$\hat{\rm o}$le of the square of the norm of ${\cal P}_k$ and it is easily computed in terms of the factors of the recurrence \cite{Finkel}:
\[
\nu_k\, =\, \prod_{j=1}^k j\epsilon (n-j+1)(\gamma+j-1)\, =\, \frac{k!n!(\gamma)_k\, \epsilon^n}{(n-k)!}
\]
The moment functional ${\cal L}$ is thus the integral:
\[
{\cal L}(p)\, =\, \int_{-\infty}^{\infty} \, p(q) \, d\Omega(q)\quad,\qquad p(q)\in {\mathbb C}[q]\, ,
\]
where $\Omega(q)$ is the function $\Omega(q)\, =\, \sum_{j=1}^{n+1} \, \Omega_j\, \theta(q-q_j)$, $\theta$ being the Heaviside step function. The coefficients $\Omega_j$ are obtained for each value of $n$ by solving the linear system $\sum_{j=1}^{n+1} {\cal P}_k(q_j)\, \Omega_j\, =\, \delta_{k0}$, $k=0,1,\dots, n$. The corresponding differential, $d\Omega(q)$, is thus a discrete Stieltjes measure for ${\cal L}$.

To finish this Section we show the polynomial solutions for the first two non-trivial values of $n$:

\noindent $\bullet$ $n=1$\, $\Leftrightarrow$ \, $\alpha=-\epsilon$. The two roots of the equation ${\cal P}_2(q)=0$ are:
\[
{\cal P}_2(q)\, =\, q^2- M q -\gamma \epsilon\, =\, (q-q_1)\, (q-q_2)
\]
\[
q_1= \frac{1}{2}
   \left(M -\sqrt{M^2+4 \epsilon
   \gamma}\right)\quad;\qquad  q_2=\frac{1}{2}
   \left(M + \sqrt{M^2+4 \epsilon
   \gamma}\right)\, .
   \]
Substitution of $q$ by $q_1$ and $q_2$ in (\ref{truncatedsol}) immediately provides the two polynomial solutions of (\ref{CHEQES}):
%\begin{eqnarray*}
%u_{1,1}(z)\, &=&\, z+ \frac{\epsilon+M + \sqrt{M^2+4 \epsilon
%  \gamma}}{\epsilon}\\ u_{1,2}(z)\, &=&\, z+ \frac{\epsilon+M -\sqrt{M^2+4 \epsilon
% \gamma}}{\epsilon}
%\end{eqnarray*}
\[
u_{1,1}(z) = z+ \frac{\epsilon+M + \sqrt{M^2+4 \epsilon
   \gamma}}{\epsilon}\, ,\    u_{1,2}(z) = z+ \frac{\epsilon+M -\sqrt{M^2+4 \epsilon
   \gamma}}{\epsilon}
\]
up to multiplicative constants.

\noindent $\bullet$ $n=2$ \, $\Leftrightarrow$ \, $\alpha=-2\epsilon$. The three polynomial solutions of (\ref{CHEQES}) are obtained from the three roots of ${\cal P}_3(q)$, $q_1,q_2$ and $q_3$. Therefore, the solutions of the cubic equation ${\cal P}_3(q)=0$ determine the polynomial solutions:
\[
u_{2,j}(z)\, =\, z^2+
  2 \left(1-\frac{2 (\gamma +1) q_j}{q_j^2-Mq_j-2 \gamma  \epsilon }\right)\, z+1 - \frac{4 (\gamma +1) (q_j-2 \gamma )}{q_j^2-Mq_j-2 \gamma  \epsilon}\  , \  j=1,2,3\, .
\]

\section{The quantum spectrum of two Coulombian centers}

As an application of the previous results we analyze in this section the existence of ``elementary" solutions of the quantum spectral problem of diatomic molecular ions, in the Born-Oppenheimer approximation \cite{Demkov}. The Schr\"odinger equation is separable in both 3D and 2D cases, leading to the Generalized Spheroidal Equation in the first and to the Razavy and Whittaker-Hill equations in the last case. The separability properties and the study of the potentials that admit exact and quasi-exact solvability for this physical problem have been recently analyzed by Miller and Turbiner in Reference \cite{Turbiner2}.

\subsection{The 3D case. The Generalized Spheroidal Equation as a QES problem}

The stationary Schr$\ddot{\rm o}$dinger equation of a charged particle moving in the potential of two fixed Coulombian centers reads:
\begin{equation}
\left( -\frac{1}{2} \Delta\, -\, \frac{Z_1}{r_1}\, -\, \frac{Z_2}{r_2}\right) \, \Psi\, =\, E\, \Psi \, \, ,\label{Sch2c}
\end{equation}
where $\hbar\, =\, m_e\, =\, e\, =\, 1$, $Z_1$ and $Z_2$ are the atomic numbers of the two nuclei, and $r_1$ and $r_2$ are the distances from the electron to the nuclei:
\[
r_1=\sqrt{\left(x_1-\frac{R}{2}\right)^2+x_2^2+x_3^2}\, ,\  r_2=\sqrt{\left(x_1+\frac{R}{2}\right)^2+x_2^2+x_3^2}\,\,   .
 \]
Equation (\ref{Sch2c}) admits separation of variables using spheroidal coordinates  \cite{Ponomarev}: $(\xi,\eta,\varphi)$, obtained by rotating the two-dimensional elliptic coordinates: $\xi\, =\, \frac{r_1+r_2}{R}\in(1,+\infty)$ and $\eta\, =\, \frac{r_2-r_1}{R}\in(-1,1)$, $R$ being the internuclear distance, about the focal axis $x_1$.

The search for wave functions of separated form:
\[
\Psi(\xi,\eta,\varphi)\, =\, F(\xi)\, G(\eta)\, e^{im\varphi}\quad,\qquad m\in {\mathbb Z}\, ,
\]
converts this PDE into the ``radial" ODE for the $\xi$ variable:
\begin{equation}
\frac{d}{d\xi} \left( (\xi^2-1) \frac{dF(\xi)}{d\xi}\right) + \left( \frac{R^2E}{2}(\xi^2-1)+R (Z_1+Z_2)\xi+\lambda -\frac{m^2}{\xi^2-1}\right)F(\xi) = 0\label{radial}
\end{equation}
and the ``angular" ODE for $\eta$:
\begin{equation}
\frac{d}{d\eta} \left( (1-\eta^2)\frac{dG(\eta)}{d\eta}\right) + \left( \frac{R^2E}{2}(1-\eta^2)+R (Z_1-Z_2)\eta-\lambda -\frac{m^2}{1-\eta^2}\right)G(\eta) = 0\label{angular}\, ,
\end{equation}
with the common separation constants $m\in\mathbb{Z}$ and $\lambda\in\mathbb{R}$. Both equations (\ref{radial}) and (\ref{angular}) are (almost identical) GSEq equations; the intervals of definition are different, $(1,+\infty)$ and $(-1,1)$, and the parameters differ in the relative sign between $Z_1$ and $Z_2$.

For this reason, solving the two equations simultaneously is no easy task. In standard approaches the integration constants $E$, $\lambda$ and $m$ provide three quantum numbers \cite{Ponomarev}. The procedure envisaged by Demkov in \cite{Demkov} is to search for eigen-wave functions of the Hamiltonian corresponding to the energy levels of a hydrogenoid atom. This bold idea demands the search for solutions of equation (\ref{radial}) with an energy of the form:
\begin{equation}
E\, =\, -\frac{(Z_1+Z_2)^2}{2n_1^2}\quad,\qquad n_1\, =\, 1,2,3, \dots \label{radialenergy}
\end{equation}
and simultaneously finding solutions of the angular equation (\ref{angular}) with energy:
\begin{equation}
E\, =\, -\frac{(Z_1-Z_2)^2}{2n_2^2}\quad,\qquad \, n_2\, =\, 1,2,3, \dots \label{angularenergy} \, \, .
\end{equation}
These two hypotheses are compatible only if the two energies coincide, and thus, the nuclear charges being fixed, there exist solutions only for values of $n_1$ and $n_2$ that solve the diophantine equation:
\begin{equation}
\frac{(Z_1+Z_2)^2}{n_1^2}=\frac{(Z_1-Z_2)^2}{n_2^2}\, \, . \label{diophantine}
\end{equation}
Moreover, the value of the separation constant $\lambda$ in (\ref{radial}) and (\ref{angular}) to be determined in the resolution process must be the same. Because of the dependence of $\lambda$ on $R$ the equality will only be satisfied for certain values of the internuclear distance.

The success of this approach relies critically on the QES character of the GSEq equation.

Synthesizing the different possibilities analyzed in Section 2, GSEq can be understood as the Confluent Heun Equation in eight non-equivalent ways, due to sign ambiguities, by using one of the two changes of variables:
\begin{eqnarray*}
v(z)\, &=&\, (z^2-1)^{ s_1\, \frac{1-\delta}{2}}\, e^{s_2\, \frac{|\epsilon|}{4} z}\, u(z) \qquad , \\ v(z)\, &=&\, \left(\frac{z-1}{z+1}\right)^{s_1 \, \frac{1-\delta}{2}}\, e^{s_2\, \frac{|\epsilon|}{4} z}\, u(z) \qquad , \qquad s_1=\pm 1\  ,\  s_2=\pm 1 \, \, .
\end{eqnarray*}
The first type of change works, i.e., it leads from the CHEq to the GSEq if $\gamma=\delta$, whereas the second one is fine when $\gamma=2-\delta$. Searching only for non-singular solutions of the GSEq with good behaviour at infinity, and assuming $\delta\geq 1$, only one sign combination is possible: $s_1=-1$, $s_2=-1$ (it is possible to develop a similar analysis for the case $\delta<1$). Thus, also requiring regularity at the $z=-1$ point we are left with the single choice:
\begin{equation}
v(z)\, =\, (z^2-1)^{ \, \frac{\delta-1}{2}}\, e^{-\, \frac{|\epsilon|}{4} z}\, u(z)\label{goodchange}\, .
\end{equation}
Applying these considerations to (\ref{radial}) a change of variable of the type (\ref{goodchange}):
\[
F(\xi)\, =\, (\xi^2-1)^{\frac{|m|}{2}}\, e^{-R(Z_1+Z_2)\,\frac{\xi}{2n_1}}\, u^r(\xi)
\]
allows us to write the radial equation (\ref{radial}) for an energy of the form (\ref{radialenergy}) as the
CHEq equation with parameters:
\begin{eqnarray*}
\gamma&=&\delta=|m|+1\, ,\quad \epsilon\, =\, -\frac{2R(Z_1+Z_2)}{n_1}\, ,\quad \alpha\, =\, 2R(Z_1+Z_2)\, \frac{n_1-|m|-1}{n_1}\\
q&=&-\lambda+R(Z_1+Z_2) \frac{n_1-|m|-1}{n_1}-|m|(|m|+1)\, .
\end{eqnarray*}
Looking at the expressions of $\alpha$ and $\epsilon$, it is clear that the Demkov hypothesis about solutions to the radial equation necessarily forces the relation $\alpha\, =\, -\, (n_1-|m|-1)\, \epsilon$. This is equivalent to the QES condition (\ref{QEScondition}) of the CHEq for the non-negative integer $n^r=n_1-|m|-1$, which requires the inequality $|m| \leq n_1-1$.

Simili modo, after the change of variable (\ref{goodchange}):
\[
G(\eta)\, =\, (1-\eta^2)^{\frac{|m|}{2}}\, e^{-R(Z_2-Z_1)\,\frac{\eta}{2n_2}}\, u^a(\eta)
\]
the angular equation (\ref{angular}) is exactly the CHEq with parameters:
\begin{eqnarray*}
\gamma&=&\delta=|m|+1\, ,\  \epsilon\, =\, -\frac{2R(Z_2-Z_1)}{n_2}\, ,\   \alpha\, =\, 2R(Z_2-Z_1)\, \frac{n_2-|m|-1}{n_2} \\
q&=&-\lambda+R(Z_2-Z_1) \frac{n_2-|m|-1}{n_2}-|m|(|m|+1)\, .
\end{eqnarray*}
Thus, the Demkov hypothesis also implies the quasi-exact solvability of the angular equation through satisfaction in the associated CHEq of $\alpha=-(n_2-|m|-1)\varepsilon$. There is another positive integer, $n^a=n_2-|m|-1$, guaranteeing the QES character of the angular equation.

In sum, the Demkov wave functions are eigenfunctions of the Hamiltonian of the form:
\begin{equation*}
\Psi(\xi,\eta,\varphi) \propto (\xi^2-1)^\frac{|m|}{2}(1-\eta^2)^\frac{|m|}{2}\, u^r(\xi)\, u^a(\eta)\, e^{-\frac{R(Z_1+Z_2)}{2n_1}\xi}e^{-\frac{R(Z_2-Z_1)}{2n_2}\eta} \, e^{im\varphi}
\end{equation*}
where $u^r(\xi)$ and $u^a(\eta)$ are respectively finite solutions of the QES radial and angular equations; i.e., polynomials up to $n_1-|m|-1$ and $n_2-|m|-1$ order.

\subsection{The 2D case. The Razavy and Whittaker-Hill Equations as QES problems}

The physical problem of two fixed Coulombian centers restricted to the two dimensional space obeys equation (\ref{Sch2c}) for $r_1=\sqrt{\left(x_1-\frac{R}{2}\right)^2+x_2^2}$ and $r_2=\sqrt{\left(x_1+\frac{R}{2}\right)^2+x_2^2}$ in the $x_1x_2$ plane. $\Delta=\frac{\partial^2}{\partial x_1^2}+\frac{\partial^2}{\partial x_2^2}$ is now the two-dimensional Laplacian. Separability in elliptic coordinates $(\xi,\eta)$ is guaranteed and the search of separated wave functions $\Psi(\xi,\eta)\, =\, F(\xi)\, G(\eta)$ converts equation (\ref{Sch2c}) into two ODEs:
\begin{equation}
(\xi^2-1)\frac{d^2F(\xi)}{d\xi^2}+ \xi \frac{dF(\xi)}{d\xi}+ \left( \frac{ER^2}{2} \xi^2+R(Z_1+Z_2) \xi+\lambda\right) F(\xi) = 0\label{radial2d}
\end{equation}
for the ``radial" coordinate $\xi\in (1,\infty)$, and
\begin{equation}
(1-\eta^2)\frac{d^2G(\eta)}{d\eta^2}- \eta \frac{dG(\eta)}{d\eta}- \left( \frac{ER^2}{2} \eta^2+R(Z_2-Z_1) \eta+\lambda\right) G(\eta) = 0\label{angular2d}
\end{equation}
for the ``angular" $\eta\in (-1,1)$ coordinate. $\lambda$ is again the separation constant.

Equations (\ref{radial2d}) and (\ref{angular2d}) are no more than equation (\ref{RWH1}), i.e. the algebraic form of Razavy and Whittaker-Hill equations respectively. Both equations are particular cases of CHEq with the changes of variable:
\begin{eqnarray*}
F(\xi)&=& (\xi+1)^{\frac{2\gamma-1}{4}}\, (\xi-1)^{\frac{2\delta-1}{4}}\, e^{\frac{\epsilon \xi}{4}}\, u(\xi)\\
G(\eta)&=&(1+\eta)^{\frac{2\gamma-1}{4}}\, (1-\eta)^{\frac{2\delta-1}{4}}\, e^{\frac{\epsilon \eta}{4}}\, u(\eta)
\end{eqnarray*}
that are compatible with the form of Razavy and Whittaker-Hill equations only in the four different cases identified in Section 2, i.e.: a) $\delta=\gamma=\frac{1}{2}$, b) $\delta=\gamma=\frac{3}{2}$, c) $\delta=\frac{1}{2},\  \gamma=\frac{3}{2}$, and d) $\delta=\frac{3}{2},\  \gamma=\frac{1}{2}$. The relation between the rest of CHEq constants and the physical parameters is determined by the identities: $
\frac{\alpha}{2}-\frac{\epsilon}{4}(\delta+\gamma)\, =\, R(Z_1+Z_2)$ in the radial equation, $\frac{\alpha}{2}-\frac{\epsilon}{4}(\delta+\gamma)\, =\, R(Z_2-Z_1)$ in the angular one, and finally: $\epsilon^2\, =\, -8ER^2$ in both equations.

The QES property for CHEq: $-\frac{\alpha}{\epsilon}\, \in {\mathbb N}$, is equivalent again to the hypothesis of hydrogenoid-type energy levels, as occurs in the 3D (Demkov) situation. Nevertheless this equivalence has now to be constructed studying independently the above mentioned cases. In the radial equation (\ref{radial2d}) the situation, for each possibility, is the following:

\begin{itemize}

\item a) $\delta=\gamma=\frac{1}{2}$. The QES condition $\alpha=-n^r \epsilon$ forces the energy and the change of variable from (\ref{radial2d}) to CHEq to be:
\[
E\, =\, \frac{-2(Z_1+Z_2)^2}{(2n^r+1)^2}\quad ;\qquad
F(\xi)\, =\, e^{-R\frac{Z_1+Z_2}{2n^r+1}\, \xi}\, u(\xi) \   .
\]

\item b) $\delta=\gamma=\frac{3}{2}$. In this case $\alpha=-n^r \epsilon$ leads to
\[
E\, =\, \frac{-2(Z_1+Z_2)^2}{(2n^r+3)^2}\quad;\qquad
F(\xi)\, =\, \sqrt{\xi^2-1}\,  e^{-R\frac{Z_1+Z_2}{2n^r+3}\, \xi}\, u(\xi)\  .
\]

\item c) $\delta=\frac{1}{2}$, $\gamma=\frac{3}{2}$. Analogously:
\[
E\, =\, \frac{-2(Z_1+Z_2)^2}{(2n^r+2)^2}\quad;\qquad
F(\xi)\, =\, \sqrt{\xi+1}\,  e^{-R\frac{Z_1+Z_2}{2n^r+2}\, \xi}\, u(\xi)\   .
\]

\item d) $\delta=\frac{3}{2}$, $\gamma=\frac{1}{2}$. Finally:
\[
E\, =\, \frac{-2(Z_1+Z_2)^2}{(2n^r+2)^2}\quad;\qquad
F(\xi)\, =\, \sqrt{\xi-1}\,  e^{-R\frac{Z_1+Z_2}{2n^r+2}\, \xi}\, u(\xi)\   .
\]

\end{itemize}
Thus the hydrogenoid-type energy \cite{GGTCM}: $E\, =\, \frac{-2 (Z_1+Z_2)^2}{n_1^2}$ is obtained in the radial equation in a different way depending on the even or odd character of $n_1$, by two pairs of changes of variable respectively.

In the angular equation (\ref{angular2d}) we find a similar situation:

\begin{itemize}

\item a) $\delta=\gamma=\frac{1}{2}$. The QES condition $\alpha=-n^a \epsilon$ now determines:
\[
E=\frac{-2(Z_1-Z_2)^2}{(2n^a+1)^2}\quad   ;\qquad G(\eta)\, =\, e^{R \frac{Z_1-Z_2}{2 n^a+1}\eta}\, u(\eta)\   .
\]
\item b) $\delta=\gamma=\frac{3}{2}$. In this case $\alpha=-n^a \epsilon$ leads to
\[
E=\frac{-2 (Z_1-Z_2)^2}{(2 n^a+3)^2}\quad   ;\qquad G(\eta)\, =\, \sqrt{1-\eta^2}\, e^{R \frac{
   Z_1-Z_2}{2 n^a+3} \eta} \,  u(\eta)\   .
\]
\item c) $\delta=\frac{1}{2}$, $\gamma=\frac{3}{2}$. Analogously:
\[
E=\frac{-2 (Z_1-Z_2)^2}{(2 n^a + 2)^2}\quad   ;\qquad G(\eta)\, =\, \sqrt{1+\eta}\,  e^{R \frac{
   Z_1-Z_2}{2 n^a+2} \eta} \, u(\eta)  \   .
\]
\item d) $\delta=\frac{3}{2}$, $\gamma=\frac{1}{2}$. Finally:
\[
E=\frac{-2 (Z_1-Z_2)^2}{(2 n^a+2)^2}\   ;\qquad G(\eta)\, =\, \sqrt{1-\eta}\,  e^{R \frac{
   Z_1-Z_2}{2 n^a+2} \eta} \, u(\eta)   \   .
\]

\end{itemize}
The same structure is thus achieved for an energy: $E\, =\, \frac{-2 (Z_1-Z_2)^2}{n_2^2}$.

Forcing simultaneous quasi-exact solvability in the two equations, for fixed values of $Z_1$ and $Z_2$, requires the verification of the same diophantine equation (\ref{diophantine}) as in the 3D case.

\subsection{Demkov wave functions}

We finally address several arrangements of charges and distances allowing the elementary solutions described by Demkov in \cite{Demkov} and we present the corresponding wave functions for the same charges in the 2D case.

\noindent $\bullet$ \  $Z_1=5$, $Z_2=1$.

We shall analyze the solution $n_1=3$, $n_2=2$ of the diophantine equation

\begin{itemize}

\item 3D Case. We shall search for finite solutions of the radial equation with $n^r=2$ and of the angular equation if $n^a=1$, assuming that $m$ is null (from a mathematical point of view it is also valid to have $m=1$ for these particular solutions of (\ref{diophantine}), but this possibility is ruled out because the corresponding solution occurs for $R=0$). Note that in this case $E=-2$.

The pertinent function and parameters for the radial equation are $F(\xi)\, =\,  e^{-R\, \xi}\, u^r(\xi)$, $\gamma=\delta=1$, $\alpha\, =\, -2\, \epsilon\, =\,  8R$, $q=4R -\lambda$. The three quadratic polynomial solutions spanning the invariant module have the form
\[
u_{2,j}^r(\xi)\, =\, \xi^2\, +\,  \left(2-\frac{8 q_j}{(q_j-2) (q_j-4 R)}\right)\, \xi\, +1-\frac{8}{q_j-4 R}\, \, ,
\]
$j=1,2,3 $, where $q_j$ are the roots of the cubic equation ${\cal P}_3(q)=0$. It is convenient to express ${\cal P}(q)$ determined from the recurrence relation (\ref{recurrence}) in terms of $\lambda=4R-q$ in such a way that the cubic equation reads:
\begin{equation}
-\lambda ^3-8 \lambda ^2+\lambda  \left(16 R^2-12\right)+64 R^2=0\label{cheq2} \quad .
\end{equation}
If we denote: $ \theta= \arccos \frac{2 \, (18 R^2-5)}{(12 R^2+7)^{3/2}}$, the standard Cardano formulas provide the three real roots:
\begin{eqnarray*}
\lambda_1=4R-q_1&=& \frac{4}{3} \left(\sqrt{12 R^2+7} \, \cos
   \frac{\theta}{3}-2\right)\\ \lambda_2=4R-q_2&=&\frac{2}{3} \left(\sqrt{12
   R^2+7} \, \left( -\sqrt{3} \sin \frac{\theta}{3}-\cos
   \frac{\theta}{3}\right)-4\right)\\ \lambda_3=4R-q_3&=&  \frac{2}{3} \left(
   \sqrt{12 R^2+7} \, \left(\sqrt{3} \sin \frac{\theta}{3}-\cos
  \frac{\theta}{3}\right)-4\right)\, .
\end{eqnarray*}
The function and parameters for the angular equation are $G(\eta)\, =\, e^{R\,\eta}\, u^a(\eta)$, $\gamma=\delta=1$, $\alpha\, =\, -\epsilon\, =\,  -4R$, $q=-2R -\lambda$. Again following the procedure explained in Section 4, two linear polynomial solutions are found:
\[
u_{1,1}^a(\eta)\, =\, \eta+\frac{1+\sqrt{4R^2+1}}{2R}\, ;\quad u_{1,2}^a(\eta)\, =\, \eta+ \frac{1-\sqrt{4R^2+1}}{2R}\, ,
\]
where $u_{1,1}^a(\eta)$ is due to the lowest root $q_1=1-2R-\sqrt{4R^2+1}$, of ${\cal P}_2(q)$, whereas $u_{1,2}^a(\eta)$ corresponds to the highest one: $q_2=1-2R+\sqrt{4R^2+1}$. The quadratic equation ${\cal P}_2(q)=0$ in terms of the separation constant $\lambda=-2R-q$, becomes:
\begin{equation}
\lambda^2\, +\, 2 \lambda-4R^2\, =\, 0\, \label{cheq1}\, .
\end{equation}

There are three quadratic polynomial solutions of the radial equation and two linear polynomial solutions of the angular equation. Both sets of solutions are compatible if the values of $\lambda$, henceforth of $R$, are identical. This is a strong condition that requires the simultaneous verification of (\ref{cheq2})  and (\ref{cheq1}). There are only four pairs of values that satisfy these two equations simultaneously: $(\lambda,R)=(0,0)$, $(\lambda,R)=(-2,0)$, $(\lambda,R)=\left( \frac{-10}{3},\frac{-\sqrt{10}}{3}\right)$, and $(\lambda,R)=\left( \frac{-10}{3},\frac{\sqrt{10}}{3}\right)$. Only the last pair makes physical sense because $R$ is positive in this case.

Between the solutions of the angular equation this last admissible pair is only compatible with the root $q_2=\frac{2}{3} \left(5-\sqrt{10}\right)$, which is tantamount to $\lambda_2=-\frac{10}{3}$. The physical value $\lambda=-\frac{10}{3}$ also selects the third root $q_3=\frac{2}{3} \left(5+2 \sqrt{10}\right)$ to fix the companion solution of the radial equation because then $\lambda_3=-\frac{10}{3}$. Henceforth, $u_{2,3}^r(\xi)$ and $u_{1,2}^a(\eta)$ are compatible finite solutions of the two QES GSEq, and the Demkov wave function
\[
\Psi(\xi,\eta,\varphi)\, =\, \left( \xi^2-2 \sqrt{\frac{2}{5}}\,  \xi-\frac{7}{5}\right) \, \left( \eta-\sqrt{\frac{2}{5}}\right) \, e^{\frac{-\sqrt{10}}{3}\, (\xi-\eta)}
\]
is an elementary eigenfunction of the Hamiltonian of two Coulombian centers with $E=-2$ and $\lambda=-\frac{10}{3}$ only when the intercenter distance is: $R=\frac{\sqrt{10}}{3}$. In Figure \ref{figure1} several contour lines and level-surfaces of the density probability:
\begin{equation}
\rho(x_1,x_2,x_3)\, =\, \frac{1}{N^2}\, | \Psi(x_1,x_2,x_3) |^2\,\, ,\   N^2\, =\, \int_{\mathbb{R}^3} \left| \Psi\right| ^2\, d^3 {\bf r}\,\, , \label{Demkov3D51}
\end{equation}
are represented.

\begin{figure}
\centerline{\includegraphics[height=4cm]{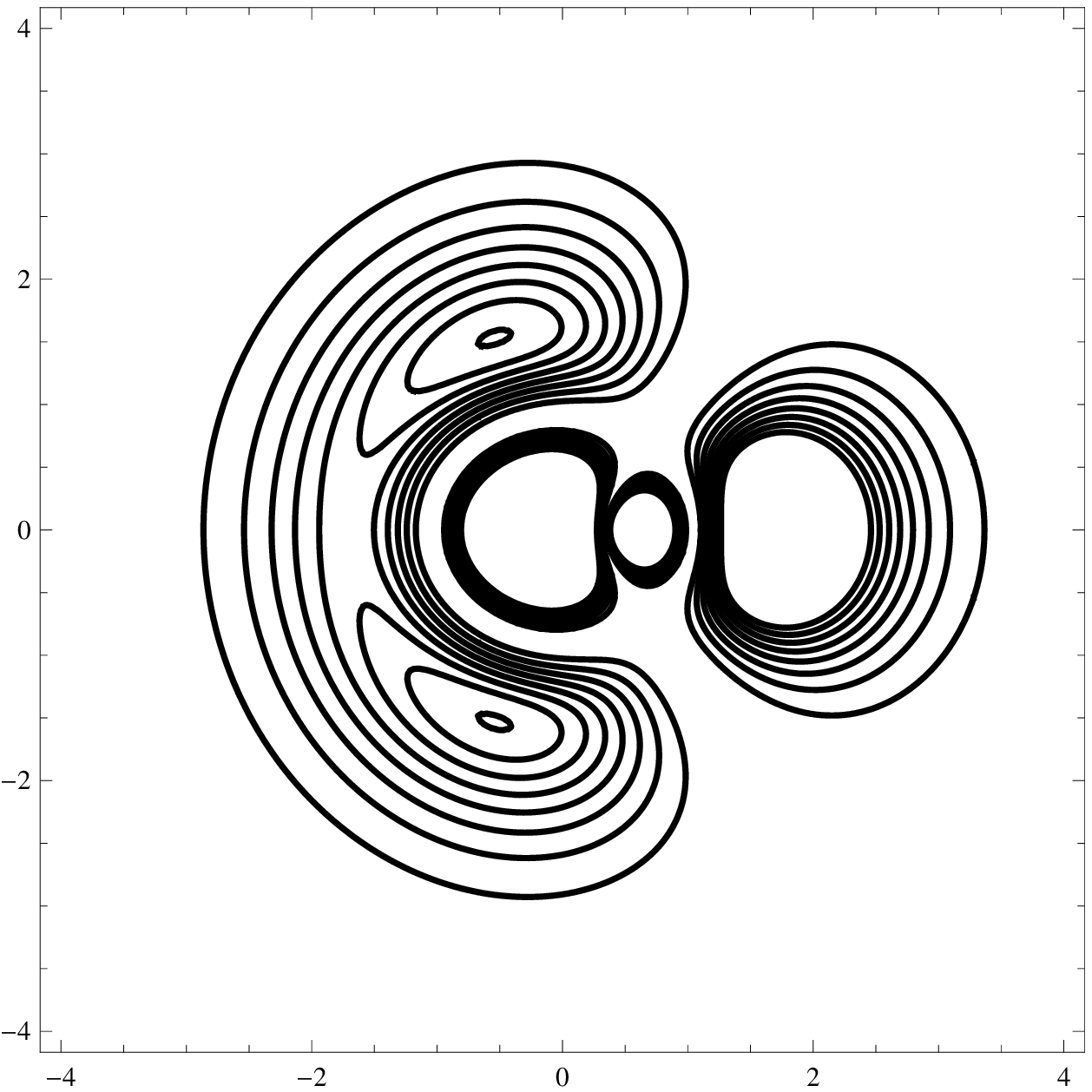}\quad \includegraphics[height=4.5cm]{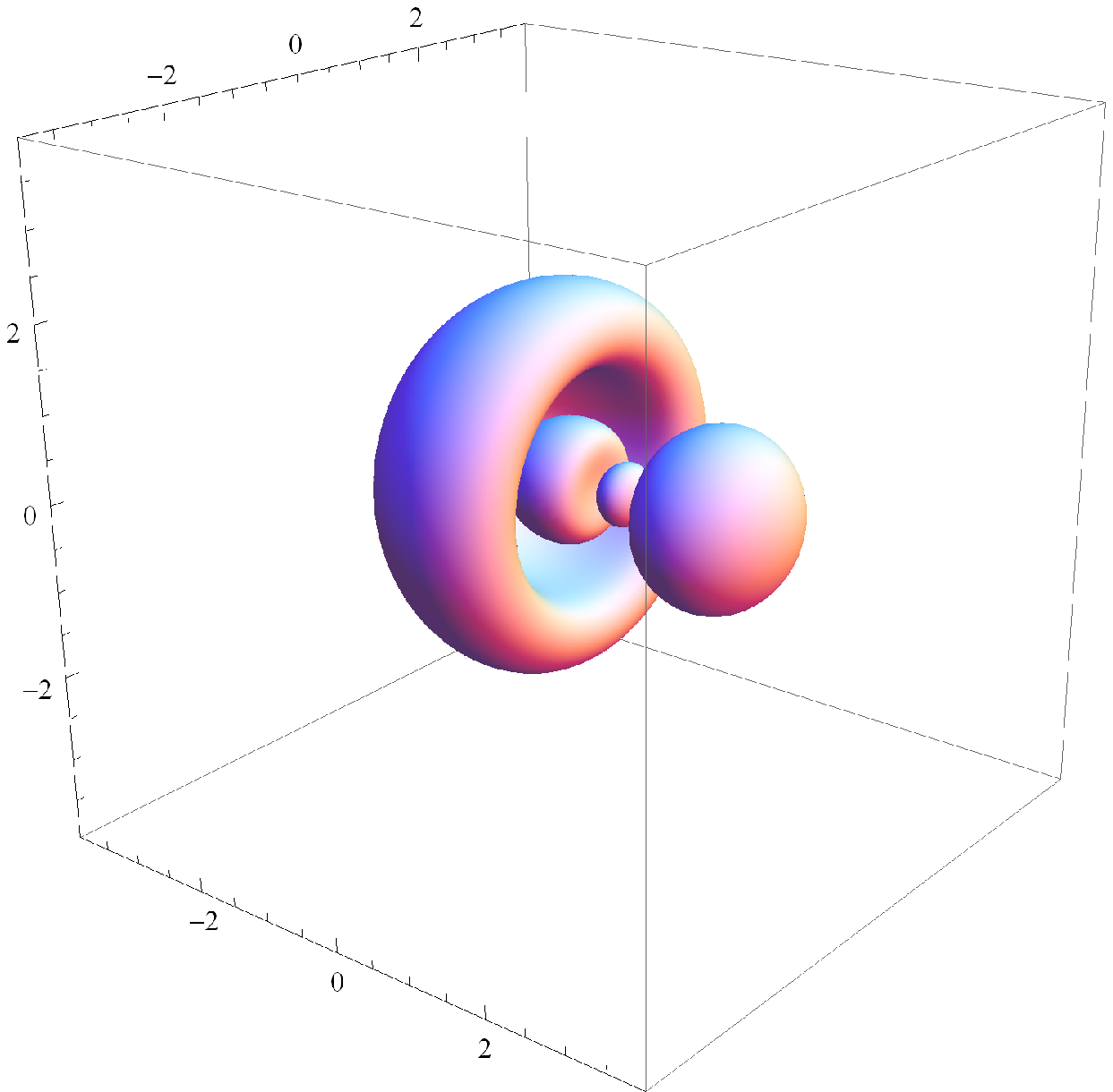}}
\caption{\label{figure1} Graphical representation of the probability density $\rho(x_1,x_2,x_3)$, (\ref{Demkov3D51}). Left) Several contour lines of $\rho(x_1,0,x_3)$, i.e. the projection of the probability density into the $x_2=0$ plane. Right) The level surface $\rho(x_1,x_2,x_3)=0.012$.}
\end{figure}

\item 2D Case. The solutions of the diophantine equation $n_1=3$ and $n_2=2$ are compatible in the Razavy equation with two possibilities: $n^r=1$ in the a)  $\delta=\gamma=\frac{1}{2}$ case, and $n^r=0$ in the b) $\delta=\gamma=\frac{3}{2}$ one. Concurrently these values of $n_1$ and $n_2$ work in the WHEq for the choices c) $\delta=\frac{1}{2}, \gamma=\frac{3}{2}$ and d) $\delta=\frac{3}{2}, \gamma=\frac{1}{2}$, in both cases with $n^a=0$.

    There are thus four systems of two algebraic equations in $\lambda$ and $R$ derived from compatibility between parameters in Razavy and WH equations. Only one solution is physically interesting, $R$ being real an positive: $(\lambda,R)=(-\frac{7}{16}, \frac{3}{8})$. These values are obtained in case b), $n^r=0$, of Razavy equation and case d), $n^a=0$, in WHEq.

    Thus we have a ``finite" planar wave function:
    \[
    \Psi(\xi,\eta)\, =\, \sqrt{\xi ^2-1} \, \sqrt{1-\eta}\, \,   e^{\frac{-3 (\xi-\eta  )}{4}}
    \]
    with: $E= -8, \lambda=-\frac{7}{16}$ and $R=\frac{3}{8}$. The corresponding probability density is given by:
    \begin{equation}
    \rho(x_1,x_2)\, =\, \frac{1}{N^2}\, \left| \Psi(x_1,x_2)\right|^2\,\, ,\   N^2\, =\, \int_{\mathbb{R}^2} \left| \Psi\right| ^2\, d^2 {\bf r}\, \, ,\label{Demkov2D51}
    \end{equation}
    see Figure \ref{figure2}.

\end{itemize}

\begin{figure}
\centerline{\includegraphics[height=4.5cm]{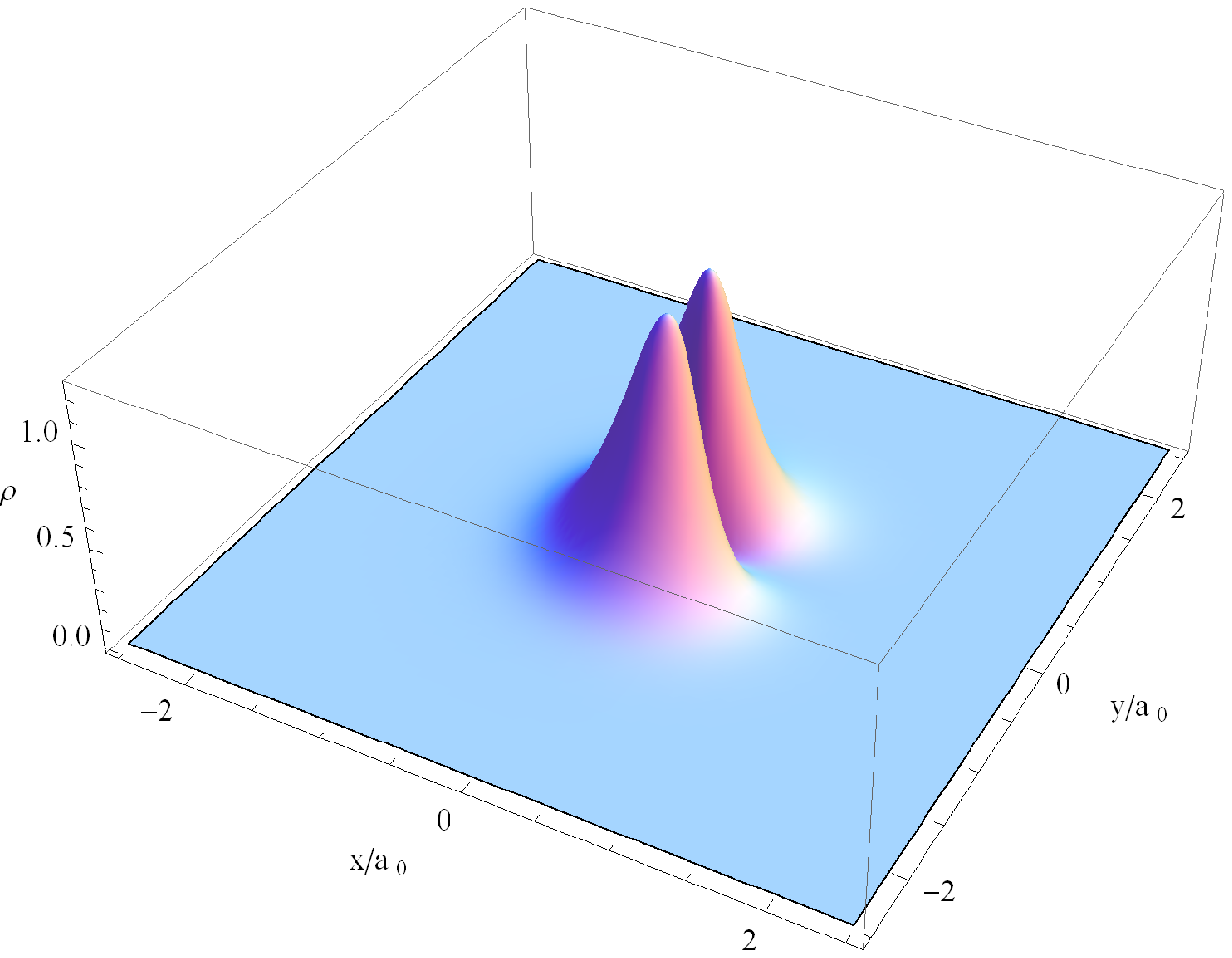}\quad \includegraphics[height=4cm]{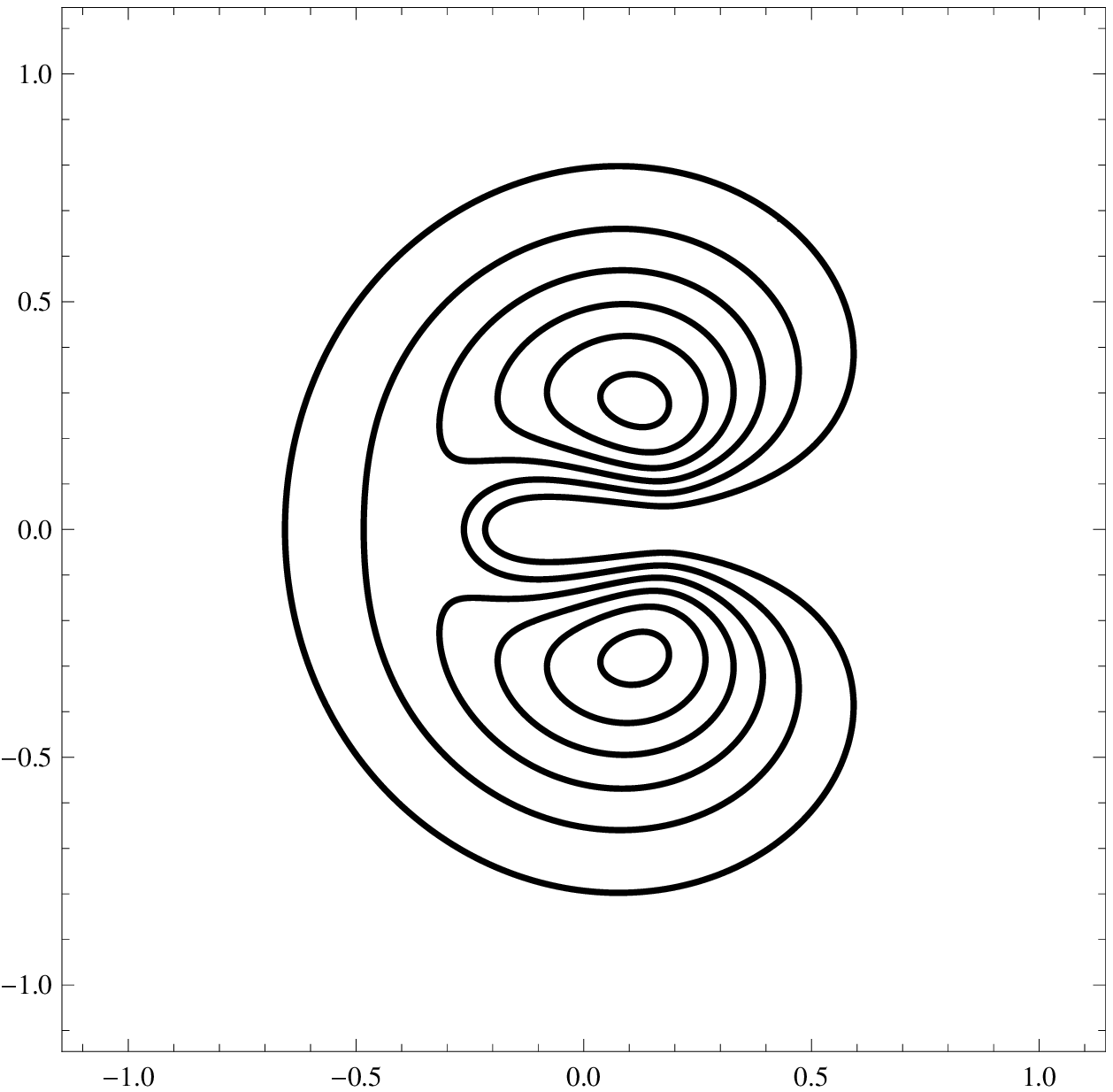}}
\caption{\label{figure2} Graphical representation of the probability density $\rho(x_1,x_2)$ in the 2D Case (\ref{Demkov2D51}). Left) 3D picture of $\rho(x_1,x_2)$. Right) Several level curves of $\rho(x_1,x_2)$.}
\end{figure}

\noindent $\bullet$ $Z_1=5$, $Z_2=3$.

We consider now the solution $n_1=4$, $n_2=1$ of the diophantine equation (\ref{diophantine}).

\begin{itemize}

\item 3D Case. $m$ is necessarily null because if not the integer $n^a$ in the angular equation would be negative. Thus, the radial and the angular equations are respectively set for $n^r=3$ and $n^a=0$. Following exactly the same procedure as in the previous case, it is found that only one of the four polynomials of third degree that solves the radial equation is compatible with the zero-degree solution of the angular equation. The internuclear distance $R$ is fixed in this case to $R=\sqrt{10}$. The Demkov wave function is simply
\[
\Psi(\xi,\eta,\varphi)\, =\,  \left( \xi^3-3 \sqrt{\frac{2}{5}}\,  \xi^2-\frac{3 }{5}\, \xi +\frac{13}{5}\,  \sqrt{\frac{2}{5}}\, \right)\, \, e^{-\sqrt{10}\, (\xi-\eta)}\, .
\]
with: $E=-2, \lambda=0$ and $R=\sqrt{10}$.

\item 2D Case. $n_1=4$ is compatible with the cases c) and d) in Razavy equation, in both cases with $n^r=1$. $n_2=1$ forces case a) with $n^a=0$ in WHEq. There merge two systems of algebraic equations in $\lambda$ and $R$, and in this case there exists two different solutions with physical interest:
\[
\Psi(\xi,\eta)\, =\,  \sqrt{\xi+1}\,  \left( \xi+4-3\sqrt{3}\right)\, \, e^{-\frac{3+2\sqrt{3}}{4} (\xi-\eta)}\, .
\]
for $E=-8, \lambda=\frac{3}{16} (7+4 \sqrt{3})$, $R=\frac{1}{8}(3+2 \sqrt{3})$, and:
\[
\Psi(\xi,\eta)\, =\,  \sqrt{\xi-1}\,  \left( \xi-4-3\sqrt{3}\right)\, \, e^{-\frac{-3+2\sqrt{3}}{4} (\xi-\eta)}\, .
\]
when $E=-8, \lambda=\frac{3}{16} (7-4 \sqrt{3})$, $R=\frac{1}{8}(-3+2 \sqrt{3})$. It is remarkable that there are two Demkov planar solutions for these values of charges while in the 3D case there is only one.

\end{itemize}

\noindent $\bullet$ $Z_1=3$, $Z_2=1$.

We focus on the quantum numbers $n_1=4$, $n_2=2$, that solve equation (\ref{diophantine}).

\begin{itemize}

\item 3D Case. Choosing the $m=0$ option the internuclear distance $R$ is fixed to $R=\sqrt{3}$. The Demkov wave function is
\[
\Psi(\xi,\eta,\varphi)\, =\, \left( \xi^3-3 \sqrt{3}\,  \xi^2+3\,  \xi + 3 \sqrt{3}\, \right)\, \left( \eta-\frac{1}{\sqrt{3}}\right) \, e^{-\frac{\sqrt{3}}{2}\, (\xi-\eta)}\, .
\]
$E=-\frac{1}{2}, \lambda= -3, R=\sqrt{3}$.

\item 2D Case. Only one solution has physical significance:
\[
\Psi(\xi,\eta)\, =\, \sqrt{\xi+1}\, \sqrt{1-\eta} \, \left( \xi-2\right) \, \, e^{-\frac{\xi-\eta}{2}}\   .
\]
corresponding to the values: $E=-2$, $\lambda=-\frac{1}{2}$, $R=\frac{1}{2}$.

\end{itemize}

Finally, it is interesting to remark in the three dimensional case that the radial equation (\ref{radial}) comes from the CHEq equation through the change of variable (\ref{goodchange}) but also implementing the second change of variables:
\begin{equation}
v(z)\, =\, \left(\frac{z-1}{z+1}\right)^{\frac{\delta-1}{2}}\, e^{-\, \frac{|\epsilon|}{4} z}\, u(z)\quad ,\label{changeextra}
\end{equation}
valid when the relation between the CHEq parameters $\gamma=2-\delta$ holds. Note that this last change of variables (\ref{changeextra}) is not applicable to the angular equation because it would be singular at $\eta=-1$.

\subsection*{Acknowledgements}

We acknowledge F. Finkel, A. Gonzalez-Lopez and M.A. Rodriguez for conversations and Seminars on their work in this area, which to a large extent constitutes the theoretical and mathematical basis of our analysis. We are also indebted to M.V. Ioffe for informing
us about the advances achieved in the physics and mathematics of two fixed centers by the Physics and Mathematics Institutes of Sankt Petersburg Federal University.


\begin{thebibliography}{10}


\bibitem{Slavyanov} Slavyanov S.Y. and Lay W., Special Functions: A Unified Theory Based on Singularities, Oxford University Press, Oxford, 2000.


\bibitem{Ronveaux} Ronveaux A., Heun's Differential Equations, Oxford University Press, Oxford, 1995.


\bibitem{Hortacsu} Hortacsu M., Heun Functions and their uses in Physics, in Proc. of the 13th Regional Conference on Mathematical Physics, Antalya, Turkey (ed. U. Camci \& I. Semiz), pp. 27--31. World Scientific, Singapore, 2013.


\bibitem{Wilson} Wilson A.H., A Generalised Spheroidal Wave Equation, \textit{Proc. Roy. Soc. A} \textbf{118} (1928) 617--635. %(DOI:10.1098/rspa.1928.0073).
Wilson A.H., The Ionised Hydrogen Molecule, \textit{Proc. Roy. Soc. A} \textbf{118} (1928) 635-647. %(DOI:10.1098/rspa.1928.0074).



\bibitem{Demkov} Demkov Y.N., Elementary Solutions of the Quantum Problem of the Motion of a Particle in the Field of Two Coulomb Centers, \textit{ZhEFT Pis'ma} \textbf{7}:3\, (1968)  101--104. English: \textit{JETP Lett.} \textbf{7}:3 \, (1968) 76--79.


\bibitem{Artemio} Finkel F., Gonzalez-Lopez A. and Rodriguez M.A., On the families of orthogonal polynomials
associated to the Razavy potential, \textit{J. Phys. A: Math. Gen.} \textbf{32} (1999) 6821--6835. %(DOI: 10.1088/0305-4470/32/39/308).

\bibitem{GGTCM} Gonzalez Leon M.A., Mateos Guilarte J., Senosiain M.J. and de la Torre Mayado M., On the Supersymmetric Spectra of two Planar Integrable
Quantum Systems, \textit{Contemporary Mathematics} \textbf{563} (2012) 73--113. %(DOI: http://dx.doi.org/10.1090/conm/563).

\bibitem{Turbiner} Turbiner A.V., Quasi-exactly-solvable problems and $sl(2)$ algebra, \textit{Commun. Math. Phys.} \textbf{118} (1988) 467--474. %(DOI: 10.1007/BF01466727).

\bibitem{Shifman} Shifman M. New Findings in Quantum Mechanics (Partial Algebraization of the Spectral Problem), in \textit{ITEP Lectures on Particle Physics and Field Theory}, vol. II, chapter VII. World Scientific Lecture Notes in Physics \textbf{62}. World Scientific, Singapore 1999.


\bibitem{Olver} Gonzalez-Lopez A., Kamran N. and Olver P.J., Normalizability of one-dimensional quasi-exactly
solvable Schr\"odinger operators, \textit{Commun. Math. Phys.}  \textbf{153} (1993) 117--146. %(DOI: 10.1007/BF02099042).

\bibitem{Leaver} Leaver E.W., An analytic representation for the quasinormal modes of Kerr black holes, \textit{Proc. R. Soc. Lond. A} \textbf{402} (1985) 285--298. %(DOI: 10.1098/rspa.1985.0119).

\bibitem{Leaver2} Leaver E.W., Solutions to a generalized spheroidal wave equation: Teukolsky's equations in general relativity, and the two-center problem in molecular quantum mechanics, \textit{J. Math. Phys.} \textbf{27} (1986) 1238--1265. %(DOI:10.1063/1.527130).

\bibitem{Figueiredo} Figueiredo B.D.B., On some solutions to generalized spheroidal wave equations and applications, \textit{J. Phys. A: Math. Gen.} {\bf 35} (2002) 2877--2906. %( DOI:10.1088/0305-4470/35/12/312).

\bibitem{Turbiner3} Turbiner A.V. and Olivares-Pilon H., The H$_2^+$ molecular ion: a solution, \textit{J. Phys. B: At. Mol. Opt. Phys.} \textbf{44} (2011) 101002 (7pp)
%doi:10.1088/0953-4075/44/10/101002


\bibitem{Turbiner2} Miller W. and Turbiner A.V., Particle in a field of two centers in prolate spheroidal coordinates: integrability and solvability, \textit{J. Phys. A: Math. Gen.} \textbf{47} (2014) 192002.

\bibitem{Ponomarev} Ponomarev L.I. and Puzynina T.P., The two-center problem in quantum mechanics, \textit{Sov. Phys. JEPT}  \textbf{25} (1967) 846--852.


\bibitem{Vincenzi} Vincenzi D. and Bodenschatz E., Single polymer dynamics in elongational flow and the confluent Heun equation, \textit{J. Phys. A: Math. Gen.} {\bf 39} (2006) 10691--10701. %(DOI:10.1088/0305-4470/39/34/007)


\bibitem{Shan} Shan Y., Jiang, T.F. and Lee Y.C., Generalized spheroidal wave equations from an image-potential method for surface effects on impurity states, \textit{Phys. Rev. B} \textbf{31} (1985) 5487--5489. %(DOI:10.1103/PhysRevB.31.5487).


\bibitem{GMT2007} Gonzalez Leon M.A., Mateos Guilarte J. and de la Torre M., Two-Dimensional Supersymmetric Quantum Mechanics: Two Fixed Centers of Force, \textit{SIGMA} {\bf 3} (2007) 124--148.

\bibitem{GMT2012} Gonzalez Leon M.A., Mateos Guilarte J. and de la Torre Mayado M., \textit{J. Phys.: Conf. Ser.} {\bf 343} (2012) 012040. %DOI:10.1088/1742-6596/343/1/012040.

\bibitem{DLMF} The Digital Library of Mathematical Functions: \S 30.12.  \textit{``Generalized and Coulomb Spheroidal Functions"}, {http://dlmf.nist.gov/30.12}, \S 30.11.  \textit{``Spheroidal Wave Functions"}, {http://dlmf.nist.gov/30.11}, \S 31.12.  \textit{``Confluent Forms of Heun's Equation"}, {http://dlmf.nist.gov/31.12}.


\bibitem{Ushveridze} Ushveridze A.G., Quasi-exactly solvable models in quantum mechanics, \textit{Sov. J. Part. Nucl.} \textbf{20} (1989) 504--528.

\bibitem{TurShifman} Shifman M. and Turbiner A.V., Quantal problems with partial algebraization of the spectrum,  \textit{Commun. Math. Phys.} \textbf{126} (1989) 347--365. %(DOI: 10.1007/BF02125129).

\bibitem{Finkel} Finkel F., Gonzalez-Lopez A. and  Rodriguez M.A., Quasi-exactly solvable potentials on the line and orthogonal polynomials, \textit{J. Math. Phys.} \textbf{37} (1996) 3954--3972. %(DOI: 10.1063/1.531591).

\end{thebibliography}
\end{document}